\documentclass[utf8]{frontiersSCNS} 

\usepackage{url,hyperref,lineno,microtype,subcaption}
\usepackage[onehalfspacing]{setspace}
\usepackage{amsmath}
\usepackage{color} 

\def\keyFont{\fontsize{8}{11}\helveticabold }
\def\firstAuthorLast{Sawicki {et~al.}} 
\def\Authors{Jakub Sawicki\,$^{1,2,*}$ and Eckehard Schöll\,$^{1,2}$}
\def\Address{$^{1}$Potsdam Institute for Climate Impact Research, Telegrafenberg A 31, 14473 Potsdam, Germany \\
$^{2}$Institut f{\"u}r Theoretische Physik, Technische Universit{\"a}t Berlin, Hardenbergstra\ss{}e 36, 10623 Berlin, Germany}
\def\corrAuthor{Jakub Sawicki}
\def\corrEmail{zergon@gmx.net}

\begin{document}
\onecolumn
\firstpage{1}

\title[Influence of sound on empirical brain networks]{Influence of sound on empirical brain networks} 

\author[\firstAuthorLast ]{\Authors} 
\address{} 
\correspondance{} 

\extraAuth{}

\maketitle

\begin{abstract}


We analyze the influence of an external sound source in a network of FitzHugh-Nagumo oscillators with empirical structural connectivity measured in healthy human subjects. We report synchronization patterns, induced by the frequency of the sound source. We show that the level of synchrony can be enhanced by choosing the frequency of the sound source and its amplitude as control parameters for synchronization patterns. We discuss a minimum model elucidating the modalities of the influence of music on the human brain.

\tiny
 \keyFont{ \section{Keywords:} synchronization, coupled oscillators, neuronal network dynamics, pattern formation: activity and anatomic, external driven} 
\end{abstract}

\section{Introduction}

Synchronization phenomena are well-known regarding dynamical activities of the brain. A high degree of synchronization is related to (slow-wave) sleep \citep{STE93b,RAT00} or transitions from wakefulness to sleep \citep{SCH08o,MOR12}. Recently, partial synchronization has become a clue to explain the first-night effect \citep{TAM16} and unihemispheric sleep \citep{RAT00,RAT16,MAS16,RAM19}. Moreover, synchronized dynamics play an important role in the dynamics of epileptic seizures \citep{GER20}, where the synchronization of a part of the brain causes dangerous consequences for the persons concerned. In contrast, synchronization is also used to explain brain processes which subserve for development of syntax and its perception \citep{KOE13,LAR15a,BAD20}. In general, synchronization theory is highly important to analyze and understand musical acoustics and music psychology \citep{JOR94,BAD13,SAW18a,HOU20,SHA20}. While the neurophysiological processes when listening to music remain ongoing research, it is presumed that a certain degree of synchrony can be observed while listening to music and building up expectations. Event-related potentials (ERPs), measured by electroencephalography (EEG) of participants while listening to music, show synchronized dynamics between different brain regions \citep{HAR14, HAR20a}. These studies indicate that the increase of synchronization represents musical large-scale form perception. Moreover, it has been observed that areas of the whole brain are involved regarding neuronal dynamics during perception \citep{BAD20}. Therefore, we propose to investigate the general influence of sound on empirical brain networks. We model the spiking dynamics of the neurons by the paradigmatic FitzHugh-Nagumo model, and investigate possible partial synchronization patterns induced by an external sound source, which is connected to the auditory cortex of the human brain. Furthermore, It is a well-known fact that an important feature of musical sound perception is tonal fusion \citep{SCH18k}. Although sound has in general a rich overtone spectrum, subjects perceive only one musical pitch which is a fusion of all partials of the spectrum. Against this background, we concentrate our general study on an external sound source with an amplitude and a single frequency, neglecting the complexity of music and its distinct effects in different frequency bands within the brain oscillations. Within the scope of this work, we have restricted ourselves to a minimal model with no node-specific behavior to reveal the impact of a periodic perturbation.

An intriguing synchronization phenomenon in networks is relay (or remote) synchronization between layers which are not directly connected, and interact via an intermediate (relay) layer \citep{LEY18}. The simplest realization of such a system is a triplex network where a relay layer in the middle acts as a transmitter between the two outer layers. Remote synchronization, a regime where pairs of nodes synchronize despite their large distances on the network graph, has been shown to depend on the network symmetries \citep{BER12,NIC13,GAM13,ZHA17,ZHA17a}. Recently the notion of relay synchronization has been extended from completely synchronized states to partial synchronization patterns in the individual layers of a three-layer multiplex network. It has been shown that the three-layer structure of the network allows for (partial) synchronization of chimera states in the outer layers via the relay layer \citep{SAW18c,SAW18,SAW20,WIN19,DRA20}. Going towards more realistic models, time-delay plays an important role in the modeling of the dynamics of complex networks. In brain networks, the communication speed will be affected by the distance between regions and therefore a stimulation applied to one region needs time to reach a different region. In such delayed system, it is possible to predict if the effects of stimulation remain focal or spread globally \citep{MUL16}. More generally, time delays due to propagation over the white-matter tracts have been shown to organize the brain network synchronization dynamics for different types of oscillatory nodes \citep{PET19}. Within the scope of this paper, we focus on the requirements for a simple model to exhibit partial synchronization patterns, which have been experimentally observed \citep{HAR14, HAR20a}. Therefore, we defer the consideration of time delays for now.

\section{Model}
\begin{figure}[tp]
    \begin{center}
\includegraphics[width = .5\textwidth]{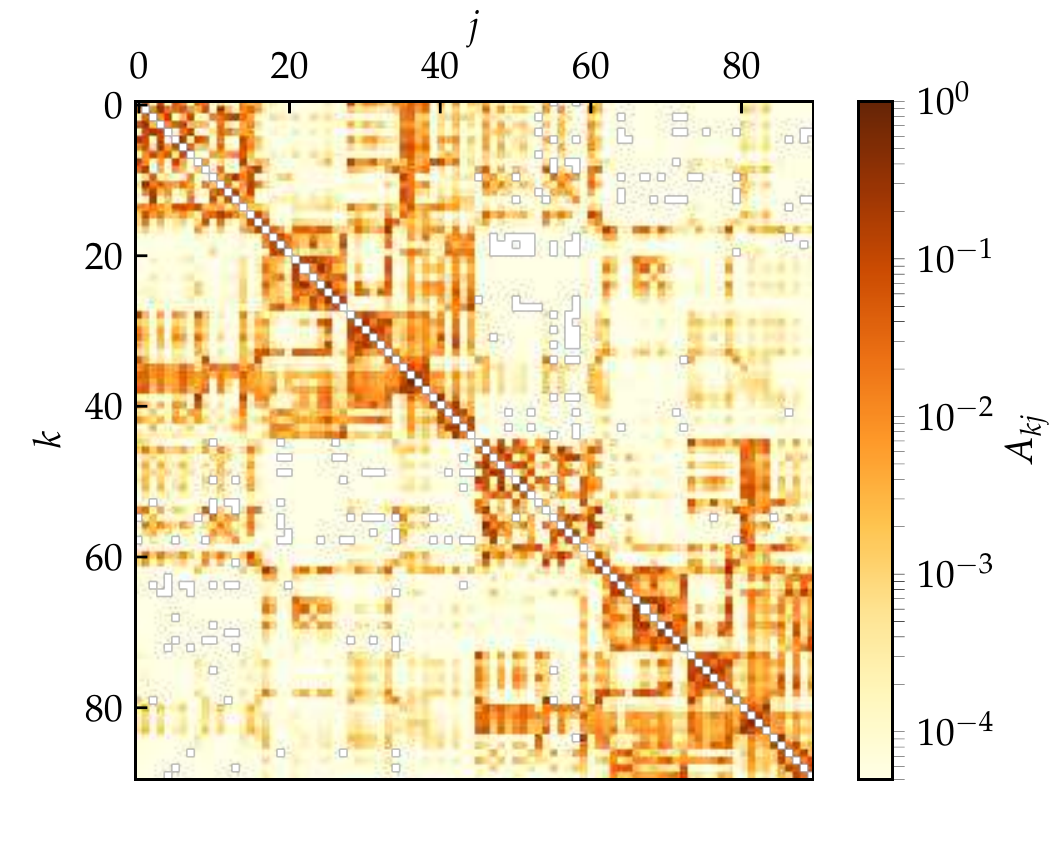}
    \caption{(color online) Model for the hemispheric brain structure: Weighted adjacency matrix $A_{kj}$ of the averaged empirical structural brain network derived from twenty healthy human subjects by averaging over the coupling between two brain regions $k$ and $j$. The brain regions $k,j$ are taken from the Automated Anatomic Labeling atlas \citep{TZO02}, but re-labeled such that $k=1,...,45$ and $k=46,...,90$ correspond to the left and right hemisphere, respectively. After \citep{GER20}.}
    \label{fig.1}
    \end{center}
\end{figure}

We consider an empirical structural brain network shown in Fig.\,\ref{fig.1} where every region of interest is modeled by a single FitzHugh-Nagumo (FHN) oscillator.
The weighted adjacency matrix $\mathbf{A} = \{A_{kj}\}$ of size $90 \times 90$, with node indices $k \in N = \{1,2,...,90\}$ was obtained from averaged diffusion-weighted magnetic resonance imaging data measured in 20 healthy human subjects. For details of the measurement procedure including acquisition parameters, see \citep{MEL15}, for previous utilization of the structural networks to analyze chimera states see \citep{CHO18,RAM19,GER20}. The data were analyzed using probabilistic tractography as implemented in the FMRIB Software Library, where FMRIB stands for Functional Magnetic Resonance Imaging of the Brain (www.fmrib.ox.ac.uk/fsl/). The anatomic network of the cortex and subcortex is measured using Diffusion Tensor Imaging (DTI) and subsequently divided into 90 predefined regions according to the Automated Anatomical Labeling (AAL) atlas \citep{TZO02}. Each node of the network corresponds to a brain region. Note that in contrast to the original AAL indexing, where sequential indices correspond to homologous brain regions, the indices in Fig.\,\ref{fig.1} are rearranged such that $k \in N_L = \{1, 2, ... ,45\}$ corresponds to left and $k \in N_R = \{46, ... ,90\}$ to the right hemisphere. Thereby the hemispheric structure of the brain, i.e., stronger intra-hemispheric coupling compared to inter-hemispheric coupling, is highlighted (Fig.\,\ref{fig.1}). 

The structural connectivity matrices serve as a realistic input for modeling, rather than as exact information concerning the existence and strength of each connection in the human brain. The pipeline for constructing such connectivity information using diffusion tractography is known to face a range of challenges~\citep{SCH19d}. While some estimates of the strength and direction of structural connections from measurements of brain activity can in principle be attempted, the relation of these can vary dramatically with (experimentally unknown) parameters of the local dynamics and coupling function~\citep{HLI12}.  

The auditory cortex is the part of the temporal lobe that processes auditory information in humans.  It is a part of the auditory system, performing basic and higher functions in hearing and is located bilaterally, roughly at the upper sides of the temporal lobes, i.e., corresponding to the AAL indexing $k = 41,86$ (temporal sup L/R). The auditory cortex takes part in the spectrotemporal analysis of the inputs passed on from the ear.

\begin{figure}
\centering
\includegraphics[width = .9\textwidth]{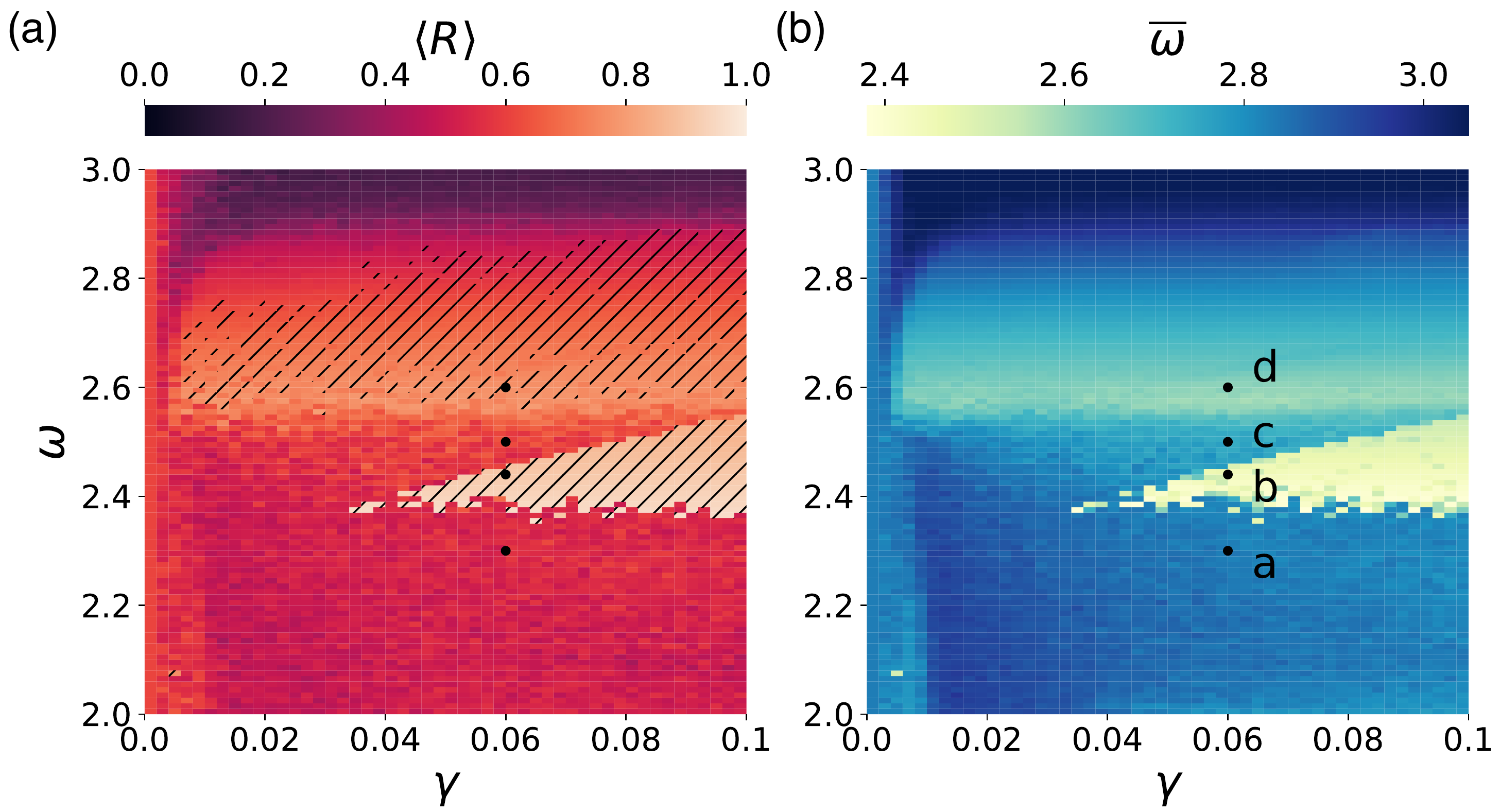}
    \caption{(color online) Synchronization tongues in brain network with external stimulus: (a) The temporal mean of the Kuramoto order parameter $\langle R \rangle$ for simulation time $\Delta T=10\,000$ and (b) the spatially averaged mean phase velocity $\overline{\omega}$ in the parameter plane of the frequency $\omega$ of the external stimulus and its amplitude $\gamma$. The light color in panel (a) stands for synchronization and the darker color for desynchronization. In the hatched region the standard deviation of $\langle R \rangle$ is less than $0.1$, which indicates the absence of strong fluctuations of $R$ in time. The dynamics of the four marked dots in each panel are shown in Figs.\,\ref{fig.3}a,b,c,d and \ref{fig.3a}a,b,c,d. Other parameters are given by $\sigma=0.6$, $\epsilon = 0.05$, $a=0.5$, and $\phi = \frac{\pi}{2} - 0.1$.}
    \label{fig.2}
\end{figure}

Each node corresponding to a brain region is modeled by the FitzHugh-Nagumo (FHN) model with external stimulus, a paradigmatic model for neuronal spiking  \citep{FIT61,NAG62,BAS18}. Note that while the FitzHugh-Nagumo model is a simplified model of a single neuron, it is also often used as a generic model for excitable media on a coarse-grained level \citep{CHE05e,CHE07a}. Thus the dynamics of the network reads:
\begin{subequations}
\begin{align}
\epsilon \dot{u}_k =& \, u_k - \frac{u_k^3}{3} - v_k \nonumber \\
                    & + \sigma \sum_{j =1}^N A_{kj} \left[ B_{uu}(u_j - u_k) + B_{uv}(v_j - v_k) \right]  \\
                     & +C_k\gamma \cos \omega t \nonumber\\
\dot{v}_k =& \, u_k + a \nonumber \\
 & + \sigma \sum_{j =1}^N  A_{kj} \left[ B_{vu}(u_j - u_k) + B_{vv}(v_j - v_k) \right], 
\end{align}
\label{eq.1}
\end{subequations}
where $\epsilon = 0.05$ describes the timescale separation between the fast activator variable (neuron membrane potential) $u$ and the slow inhibitor (recovery variable) $v$ \citep{FIT61}. Depending on the threshold parameter $a$, the FHN model may exhibit excitable behavior ($\left| a \right| > 1$) or self-sustained oscillations ($\left| a \right| < 1$). We use the FHN model in the oscillatory regime and thus fix the threshold parameter at $a=0.5$ sufficiently far from the Hopf bifurcation point. The external stimulus is modeled by a trigonometric function with frequency $\omega$ and amplitude $\gamma$ and is applied to the brain areas $k=41,86$ associated with the auditory cortex, i.e. $C_k=1$ if $k=41$ or $86$ and zero otherwise. The coupling between the single regions is given by the coupling strength $\sigma$. As we are looking for partial synchronization patterns we fix $\sigma = 0.6$ similar to numerical studies of synchronization phenomena during unihemispheric sleep \citep{RAM19} and epileptic seizures \citep{GER20} where partial synchronization patterns have been observed. The interaction scheme between nodes is characterized by a rotational coupling matrix:
\begin{equation}
\mathbf{B} = 
\begin{pmatrix}
B_{uu} & B_{uv} \\
B_{vu} & B_{vv}
\end{pmatrix}
=
\begin{pmatrix}
\text{cos}\phi & \text{sin}\phi \\
-\text{sin}\phi & \text{cos}\phi
\end{pmatrix},
\end{equation}
with coupling phase $\phi = \frac{\pi}{2} - 0.1$, causing primarily an activator-inhibitor cross-coupling. This particular scheme was shown to be crucial for the occurrence of partial synchronization patterns in ring topologies \citep{OME13} as it reduces the stability of the completely synchronized state. Also in the modeling of epileptic-seizure-related synchronization phenomena \citep{GER20}, where a part of the brain synchronizes, it turned out that such a cross-coupling is important. The subtle interplay of excitatory and inhibitory interaction is typical of the critical state at the edge of different dynamical regimes in which the brain operates \citep{MAS15a}, and gives rise to partial synchronization patterns which are not found otherwise.

\section{Methods}
We explore the dynamical behavior by calculating the mean phase velocity ${\omega_k = 2\pi M_k/\Delta T}$ for each node $k$, where $\Delta T$ denotes the time interval during which $M$ complete rotations are realized. Throughout the paper we use $\Delta T = 10\,000$. For all simulations we use initial conditions randomly distributed on the circle $u_k^2+v_k^2=4$. In case of an uncoupled system ($\sigma=0$), the mean phase velocity (or natural frequency) of each node is $\omega_k = \omega_{\text{FHN}} \approx 2.6$. Furthermore we introduce hemispheric measures that characterize the degree of synchronization of the sub-networks and give complementary information. First, the spatially averaged mean phase velocity is: 
\begin{equation}
{\overline\omega = \frac{1}{90}\sum_{k =1}^N \omega_k},    
\end{equation}
Thus $\overline\omega$ corresponds to the mean phase velocity averaged over the left and right hemisphere.
Second, the Kuramoto order parameter:
\begin{equation}
{R(t) = \frac{1}{90} \left| \sum_{k =1}^N \text{exp}[i \theta_k(t)]\right|},
\end{equation}
is calculated by means of an abstract dynamical phase $\theta_k$ that can be obtained from the standard geometric phase ${\tilde{\phi}_k(t) = \text{arctan}(v_k/u_k)}$ by a transformation which yields constant phase velocity $\dot{\theta}_k$. For an uncoupled FHN oscillator the function $t(\tilde{\phi}_k)$ is calculated numerically, assigning a value of time $0<t(\tilde{\phi}_k)<T$ for every value of the geometric phase, where $T$ is the oscillation period. The dynamical phase is then defined as $\theta_k=2 \pi t(\tilde{\phi}_k)/T$, which yields $\dot{\theta}_k = \text{const}$. Thereby identical, uncoupled oscillators have a constant phase relation with respect to the dynamical phase. Fluctuations of the order parameter $R$ caused by the FHN model's slow-fast time scales are suppressed and a change in $R$ indeed reflects a change in the degree of synchronization. The Kuramoto order parameter may vary between 0 and 1, where $R=1$ corresponds to complete phase synchronization, and small values characterize spatially desynchronized states. Additionally, we calculate the temporal mean of the Kuramoto order parameter
\begin{equation}
\langle R(t) \rangle = \frac{1}{\Delta T}\int_0^{\Delta T} R(t) \,\mathrm{d} t
\end{equation}
to estimate the general dynamical behavior of the system over time. Similarly, the temporal mean $\langle \Omega(t) \rangle $ of the collective frequency $\Omega$ of the mean field \citep{PET13}, defined by
\begin{equation}
\Omega(t)\equiv \dot{\psi}(t), \quad     R(t) e^{i \psi(t)} = \frac{1}{90} \sum_{k =1}^N \text{exp}[i \theta_k(t)]
\end{equation}
can be considered, and compared with the spatially averaged mean phase velocity.

\begin{figure}
\centering
\includegraphics[width = .85\textwidth]{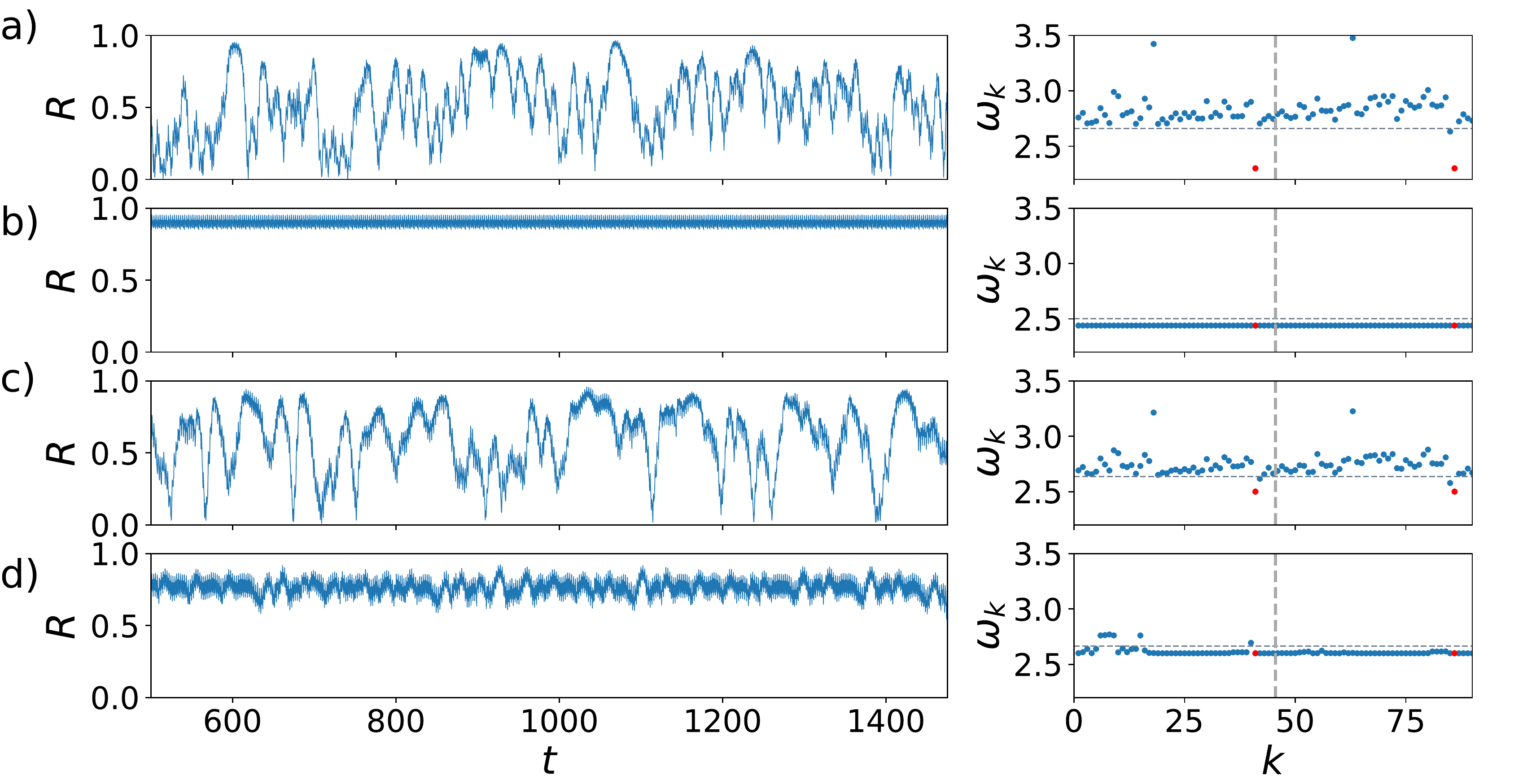}
    \caption{(color online) Dynamical scenarios: dynamics inside and outside the synchronization regions (marked as black dots in Fig.\,\ref{fig.2}) by the Kuramoto order parameter $R$ (left column) and the mean phase velocities $\omega_k$ (right column) for increasing values of the frequency $\omega$ of the external stimulus $\omega=2.30$ (a), $\omega=2.44$ (b), $\omega=2.50$ (c), and $\omega=2.60$ (d) for fixed amplitude $\gamma = 0.06$. The vertical dashed line in the right column separates the left and right hemisphere; the horizontal grey dotted line indicates the temporal average of the mean-field frequency $\Omega$. The red dots mark the nodes of the auditory cortical regions ($k=41,86$). Other parameters are as in Fig.\,\ref{fig.2}.}
    \label{fig.3}
\end{figure}

\section{Synchronization regions}

We investigate synchronization scenarios emerging from an external periodic stimulus in the auditory cortices of both hemispheres ($k=41,86$). Figure~\ref{fig.2} shows synchronization scenarios of an empirical structural brain network in dependence of the frequency $\omega$ and amplitude $\gamma$ of the external stimulus. The light colored regions in Fig.\,\ref{fig.2}a indicates synchronized dynamics, whereas the darker colors indicate desynchronized dynamics. There is a light colored stripe for $\omega=2.6$ which indicates a Kuramoto order parameter $\langle R \rangle \approx 0.8$ and a light colored tongue starting at $\omega=2.4, \gamma=0.04$. The hatched region in Fig.\,\ref{fig.2}a stands for a low standard deviation $<0.1$ of the temporal mean of the Kuramoto order parameter $\langle R \rangle$. It indicates the absence of strong fluctuations of $R(t)$ and therefore a constant high level of synchrony in time. Figure~\ref{fig.2}b shows the drop of the spatially averaged mean phase velocity $\overline{\omega}$ in case of coherent dynamics in the synchronization regions of Fig.\,\ref{fig.2}a. In the upper region, $\overline{\omega}$ takes over the value of the frequency $\omega$ of the external stimulus, whereas in the synchronization tongue $\overline{\omega}$ keeps its value of $\overline{\omega}=2.4$. 

It turns out that by taking the frequency $\omega$ of the external stimulus as a control parameter, one can change the level of synchrony of the system. Figure \ref{fig.3} depicts the details of the transition to synchronization for increasing values of the frequency $\omega$ of the external stimulus. Fixing the amplitude $\gamma = 0.06$, we take a closer look on the temporal evolution of $R$ and the mean phase velocities in the system for different regions in Fig.\,\ref{fig.2}: In Fig.\,\ref{fig.3}a the temporal evolution of the Kuramoto order parameter is similar to the system behavior without external stimulus, i.e., it exhibits large temporal fluctuations. In the right column the phase velocities of all nodes are plotted, the horizontal grey dotted line indicates the temporal average of the collective mean-field frequency $\Omega$. Only the phase velocity of the auditory cortex follows the frequency of the external driving stimulus $\omega=2.3$ and therefore is lower than the frequency of the other nodes $\omega_k\approx 2.8$. Increasing the external frequency to $\omega=2.4$ yields an abrupt transition to a synchronized state. In Fig.\,\ref{fig.3}b the Kuramoto order parameter $R \approx 0.95$ and the mean phase velocities indicate a synchronous dynamical behavior, which agrees with the collective frequency $\Omega$ of the mean-field (grey dotted horizontal line). With a further increment to $\omega=2.5$, the system loses synchrony (see Fig.\,\ref{fig.3}c) and enters the region between the two synchronization regions in Fig.\,\ref{fig.2}a. For $\omega=2.6$ in Fig.\,\ref{fig.3}d, which corresponds to the natural frequency of the uncoupled oscillators, the system regains synchronization, though the Kuramoto order parameter with $R \approx 0.8$ is lower than in the synchronization tongue. Remarkable is the fact of a dynamical asymmetry shown by the mean phase velocities. While the nodes of the right hemisphere exhibit an equal mean phase velocity, a part of the left hemisphere exhibits a faster dynamic similar to dynamics of unihemispheric sleep studied in \citep{RAM19}. In such states one hemisphere is synchronized, whereas the other hemisphere is partly desynchronized.

\begin{figure}
\centering
\includegraphics[width = .65\textwidth]{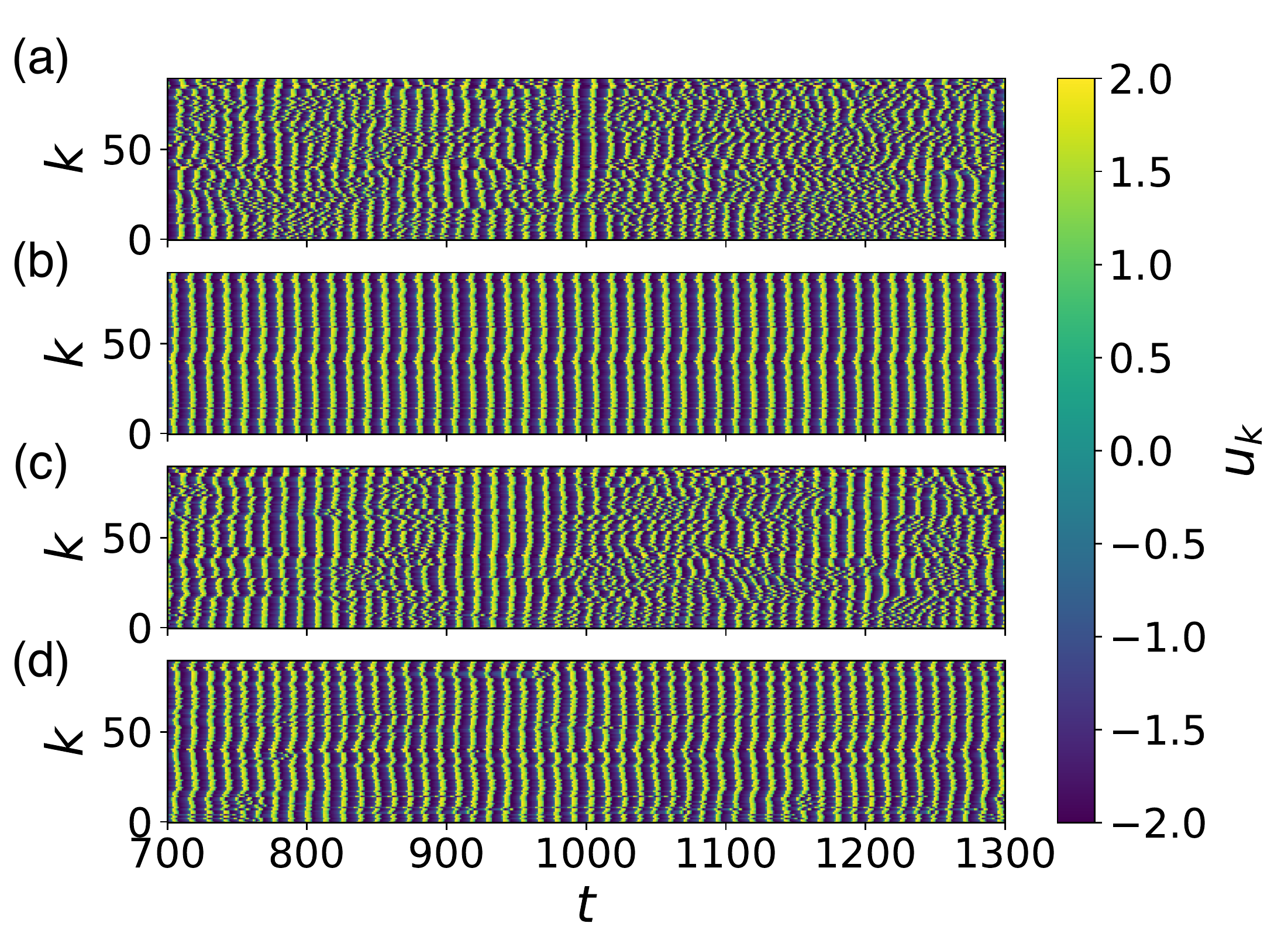}
    \caption{(color online) Synchronized and desynchronized dynamics: Shown are the space-time plot of the variable $u_k$ inside and outside the synchronization regions (marked as black dots in Fig.\,\ref{fig.2}) for increasing values of the frequency $\omega$ of the external stimulus $\omega=2.30$ (a), $\omega=2.44$ (b), $\omega=2.50$ (c), and $\omega=2.60$ (d) for fixed amplitude $\gamma = 0.06$. The panels correspond to the panels in Fig.\,\ref{fig.3}. Other parameters are as in Fig.\,\ref{fig.2}.}
    \label{fig.3a}
\end{figure}

For a better insight, Fig.\,\ref{fig.3a} shows the space-time plot of the variable $u_k$ for the corresponding parameter values in Fig.\,\ref{fig.3}. In Fig.\,\ref{fig.3a}b,d, the dynamics inside the two synchronization regions is depicted. The perturbation in the mean phase velocity profile in the right panel of Fig.\,\ref{fig.3}d, can be detected also in the corresponding perturbations in Fig.\,\ref{fig.3a}d. Comparing Fig.\,\ref{fig.3a}a and c, we can see an increase of synchronized time segments. This increase will be analyzed quantitatively in more detail in the inset of Fig.\,\ref{fig.4}.

\section{Transition to synchronization}

There are two frequencies which play an important role for the dynamics of the system. On the one hand, in Fig.\,\ref{fig.2}a a broad synchronization region is located at a frequency $\omega \approx 2.6$, which is the frequency of the uncoupled FHN oscillator $\omega_{\text{FHN}}$. Although the external stimulus effects only the two auditory nodes ($k=41,86$), we can observe a transition to synchronization of the whole system approaching $\omega\approx 2.6$ already for small values of the amplitude $\gamma > 0.004$. On the other hand, we can detect a synchronization tongue with a lower boundary at $\omega\approx 2.4$ and an upper boundary increasing linearly with the amplitude $\gamma$. In contrast to the first, smooth transition, we can find here a sharp transition to synchronized dynamics, similar to a first order transition, depicted by the high contrast of the boarder of the synchronization tongue in Fig.\,\ref{fig.2}a. In this synchronization tongue, the nodes oscillate with an equal mean phase velocity (see Fig.\,\ref{fig.3}b), but there are phase differences between them, as indicated by $0.95<R(t)< 1$ and shown in the phase-time plot in Fig.\,\ref{fig.3a}b. Using the fact that $u_j$/$v_j$ and $u_k$/$v_k$ are on the same limit cycle in the phase space and have the same mean phase velocity, the phase differences in the coupling term of Eq.\eqref{eq.1} can be effectively summed up in following way:
\begin{equation}
\sum_{j} A_{kj}\mathbf{B}\begin{pmatrix} u_j-u_k \\ v_j-v_k \end{pmatrix}\approx \Delta t_{\text{eff }}\mathbf{B}\begin{pmatrix} \dot{u}_k \\ \dot{v}_k \end{pmatrix},
\label{eq.2}
\end{equation}
where $\Delta t_{\text{eff }}\ll 1$ denotes the effective sum of the time intervals of all phase differences. Neglecting $\cos \phi \ll 1$ and setting $\sin \phi \approx 1$, Eq.\,\eqref{eq.1} reads for $k \neq 41,86$:
\begin{equation}
\begin{aligned}
\varepsilon \dot{u}_k &= u_k - \frac{u_k^3}{3} - v_k - \sigma \Delta t_{\text{eff }} \dot{v}_k\\
\dot{v}_k &= u_k + a + \sigma \Delta t_{\text{eff }} \dot{u}_k
\end{aligned}
\label{eq.3}
\end{equation}
The local dynamics of Eq.\,\eqref{eq.1} is governed by a slow-fast system (FitzHugh-Nagumo oscillator), where the slow part essentially determines the period of the oscillations. Hence, considering the slow motion on the falling branches of the $u$-nullcline ($\dot{u}_k=0$) by inserting the second equation into the first one
\begin{equation}
v_k = u_k - \frac{u_k^3}{3}-\sigma \Delta t_{\text{eff }} (u_k+a),
\end{equation}
the time derivative of the falling branches yields with $\dot{v}_k$ from Eq.\,\eqref{eq.3}
\begin{equation}
\dot{u}_k=\frac{u_k+a}{1 -u_k^2-2\sigma \Delta t_{\text{eff }}}.
\end{equation}
The separation of the variables gives
\begin{equation}
\text{d} t=\frac{1 -u_k^2-2\sigma \Delta t_{\text{eff }}}{u_k+a}\, \text{d} {u}_k,
\end{equation}
where $\text{d} t$ can be integrated over one oscillation period $T$. As shown in \citep{SAW19}, this leads in case of synchronization to a linear dependence of the oscillation period $T_{\text{sync}}= \int_0^T \text{d} t$ on the effective sum of the phase differences proportional to $\Delta t_{\text{eff }}$. For incoherent distribution of the phases of each node $k$ (see Fig.\,\ref{fig.3a}d), the phase differences between the single nodes are also strongly distributed and thus $\Delta t_{\text{eff }} \approx 0$. In this case, the natural frequency of the uncoupled system plays an important role, provided that the mean phase velocity of all oscillators is still almost equal as in case of Fig.\,\ref{fig.2}d. 

This could explain on one side the fact that we observe a synchronization tongue at $\omega\approx 2.4$ (which is smaller than the frequency of an uncoupled oscillator $\omega_{\text{FHN}} \approx 2.6$), and on the other side, the linear boundaries of the synchronization tongue for increasing amplitude $\gamma$. The increase of $\gamma$ yields an increase of the sum of the phase differences in the coupling term of Eq.\eqref{eq.1} and therefore an increase of the effective sum of the time intervals $\Delta t_{\text{eff }}$. 
\begin{figure}
\centering
\includegraphics[width=.85\linewidth]{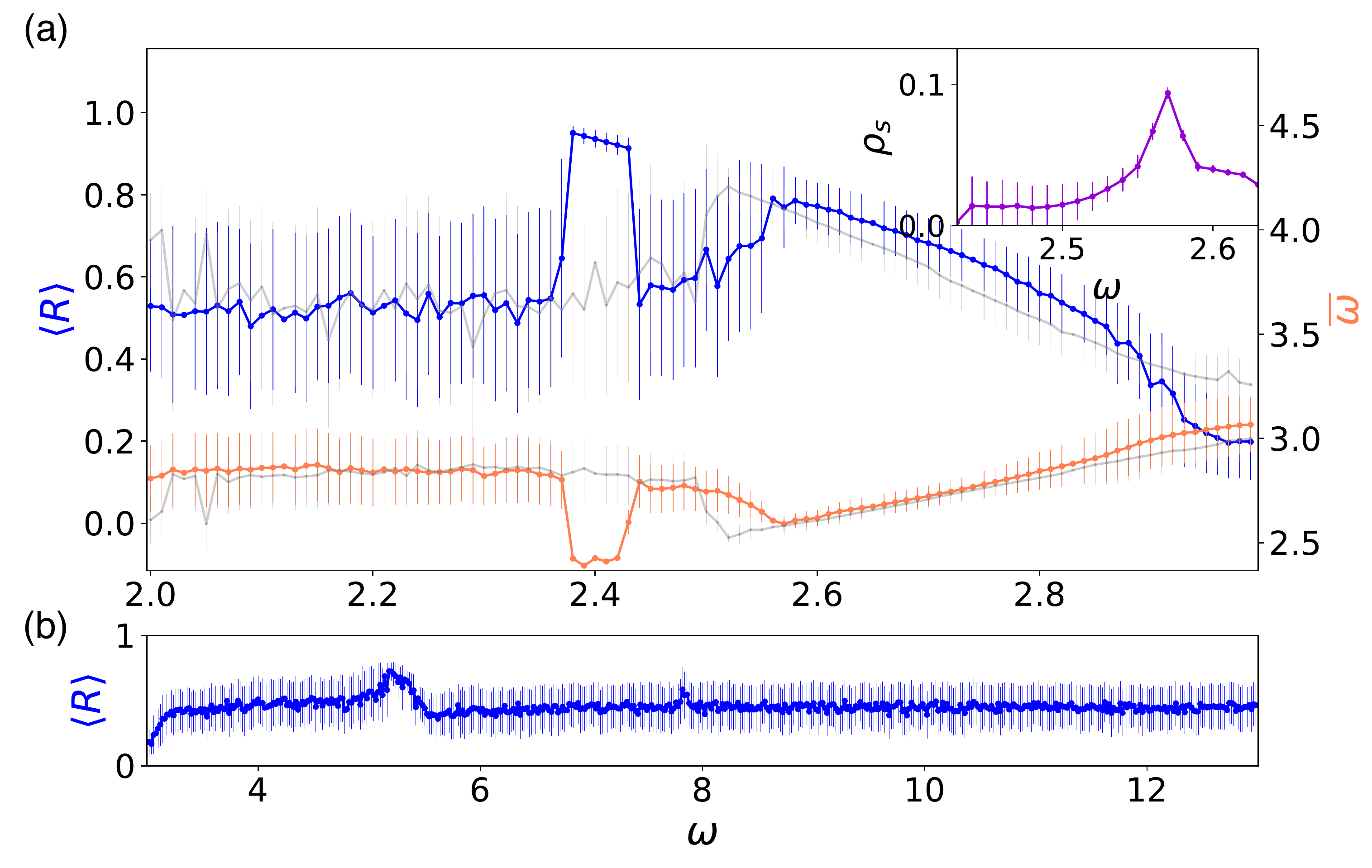}
\caption{(color online) Transition scenarios: (a) temporal mean of the Kuramoto order parameter $\langle R \rangle$ (dark blue) and the spatially averaged mean phase velocities $\overline{\omega}$ (light orange) in dependence on the frequency $\omega$ of the external stimulus for a fixed amplitude $\gamma=0.052$. The vertical bars indicate the standard deviation of the temporal mean of the Kuramoto order parameter and the spatially averaged mean phase velocities, respectively. As input nodes, the auditory cortices $k=41,86$ are chosen. In case of a different input ($k=1,45$) the corresponding light grey curves are shown in panel (a). The inset in panel (a) depicts $\rho_s=\frac{N_s}{\Delta T_L}$, the number $N_s$ of synchronized time intervals ($R(t)>0.8 \;\forall \,t$) divided by a simulation time of $\Delta T_L=30\,000$ for values of the frequency $\omega$ between the two synchronization regions. The vertical bars denote the standard deviation of the length of these synchronized time intervals. (b) $\langle R \rangle$ for a larger range of driving  frequencies $\omega$, showing higher resonance tonges. Other parameters are as in Fig.\,\ref{fig.2}.}
\label{fig.4}
\end{figure}

In Fig.\,\ref{fig.4}a, both transitions are depicted in dependence on the frequency $\omega$ for a fixed amplitude $\gamma=0.052$. We can see an abrupt increase and decrease of the temporal mean of the Kuramoto order parameter $\langle R \rangle$ before and after $\omega\approx2.4$, respectively. In contrast, in approaching the upper synchronization region starting from $\omega\approx2.6$, $\langle R \rangle$ increases more slowly than at the transition to the synchronization tongue ($\omega\approx2.4$). In case of synchronization the standard deviation of $\langle R \rangle$, displayed by the vertical bars, is smaller than in case of desynchronized dynamics. That holds also for the spatially averaged mean phase velocities $\overline{\omega}$, which in case of synchronization takes over the lower value of the frequency $\omega$ of the external stimulus. Also for $\omega>2.6$, $\overline{\omega}$ is equal to $\omega$, whereas the standard deviation of $\overline{\omega}$ increases linearly with $\omega$. In contrast, there is no effect on the system for $\omega<2.4$. Neither $\langle R \rangle$ nor $\overline{\omega}$ show a different behavior for such values of $\omega$. The high value of the standard deviation of $\langle R \rangle$ stands for dynamics as shown in Fig.\,\ref{fig.3}a, where the Kuramoto order parameter $R(t)$ is fluctuating over its whole bandwidth $R \in [0,1]$. Simulations show that for $\omega>3.0$ the dynamical behavior of the system becomes similar to that with $\omega<2.3$. For both parameter intervals of $\omega$, there is no effect on the system. Simulations show also that a similar transition to synchronization at $\omega=2.6$ can be found for higher harmonics, i.e., multiple values of $\omega=2.6$. In Fig.\,\ref{fig.4}b, we can identify synchronization regions for $\omega=5.2, 7.8$ and $10.4$ becoming less pronounced for increasing $\omega$, i.e., having a smaller extension in the plane of $\omega$ and $\gamma$. In contrast, we could not detect repeated synchronization tongues of $\omega$ for multiple values of $\omega=2.4$. This indicates the existence of two different synchronization mechanisms.  

The existence of two synchronization regions depends on the choice to which nodes the external stimulus is supplied. In case of a different input, for instance $k=1,45$ in contrast to $k=41,86$, the light grey curves in Fig.\,\ref{fig.4}a depict the corresponding dependence of the Kuramoto order parameter $\langle R \rangle$ and the spatially averaged mean phase velocities $\overline{\omega}$ upon the frequency $\omega$ of the external stimulus. The synchronization region at $\omega\approx 2.4$ is missing here and only one synchronization region remains ($\omega > 2.5$).

\begin{figure}
\centering
\includegraphics[width = .85\textwidth]{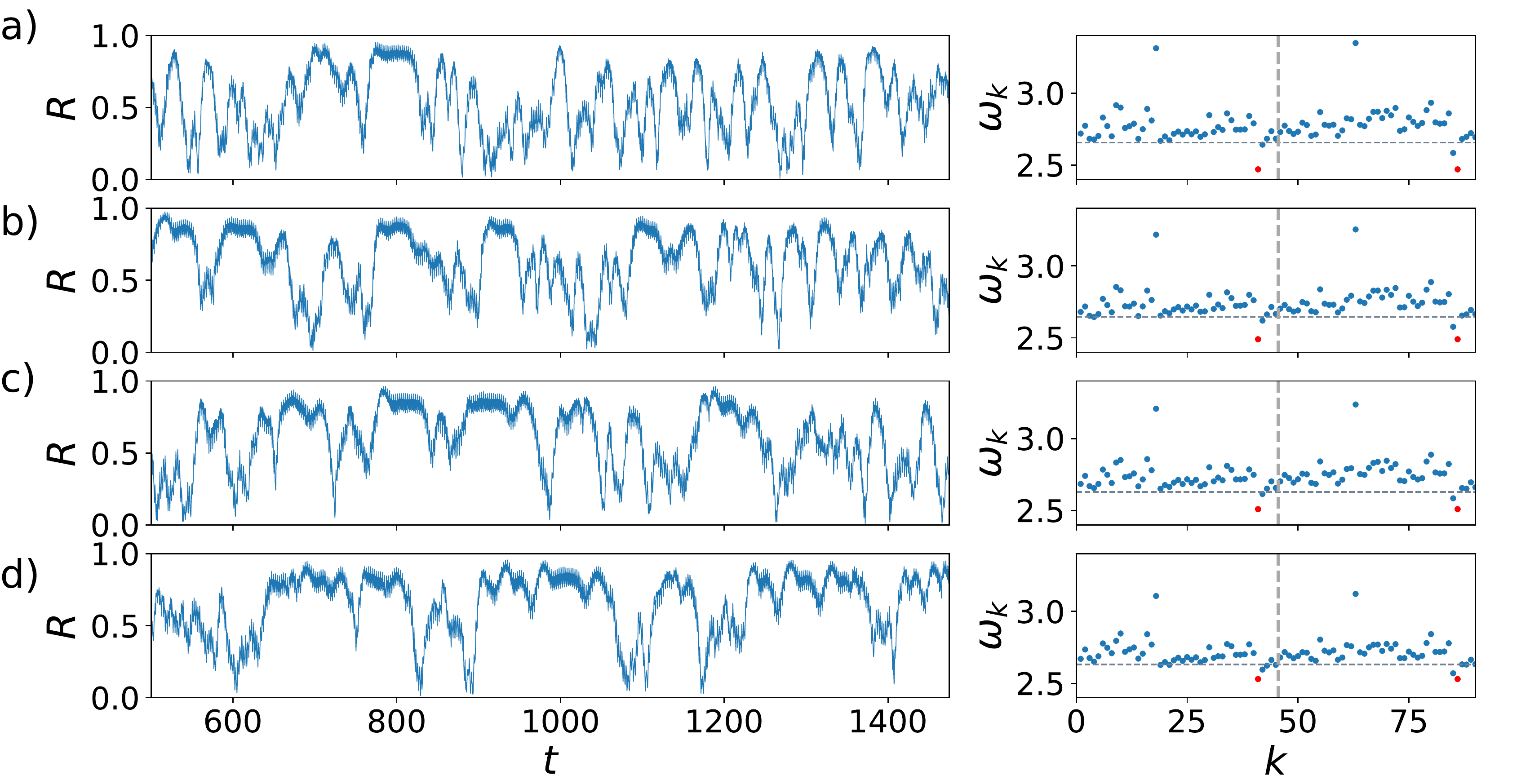}
\caption{(color online) Transition scenarios: The increase of the regularity and duration of synchronized time intervals shown by the temporal evolution of the Kuramoto order parameter $R$ (left column) and the mean phase velocities $\omega_k$ (right column) for increasing values of the frequency $\omega$ of the external stimulus $\omega=2.47$ (a), $\omega=2.49$ (b), $\omega=2.51$ (c), and $\omega=2.53$ (d) for fixed amplitude $\gamma = 0.06$. The vertical dashed line in the right column separates the left and right hemisphere, and the horizontal grey dotted line indicates the temporal average of the mean-field frequency $\Omega$. The red colored dots indicate the nodes of the auditory cortical regions ($k=41,86$). Other parameters as in Fig.\,\ref{fig.2}.}
\label{fig.5}
\end{figure}

In the following, we analyze the region between the two synchronization areas in more detail. Figure \ref{fig.5} depicts the dynamical behavior when we approach the synchronization region by increasing the frequency $\omega$ of the external stimulus in the neighborhood of the synchronization region at $\omega=2.6$. For $\omega=2.47$ in Fig.\,\ref{fig.5}a, the time series of the Kuramoto order parameter shows familiar temporal fluctuations with only short episodes of synchronization ($R(t)>0.8$). In \citep{GER20} the authors define the threshold $R=0.8$ as the onset of an epileptic seizure. By increasing the frequency $\omega$, one can increase the quantity of these episodes, as well as their duration. Figure \ref{fig.5}d with $\omega=2.51$ features much longer duration of synchronized episodes, moreover the duration of the single episodes are comparable in length. This transition in Fig.\,\ref{fig.5}a-d can be also seen in Fig.\,\ref{fig.4}a. The inset of Fig.\,\ref{fig.4}a confirms the increasing regularity between the two synchronization regions by depicting $\rho_s=\frac{N_s}{\Delta T_L}$ versus $\omega$, where $N_s$ is the number of synchronized time intervals ($R(t)>0.8 \, \forall \, t$) and $\Delta T_L=30\,000$ is the simulation time. The vertical bars denote the standard deviation of the length of these synchronized time intervals. With increasing $\omega$ not only the number of synchronized time intervals is increasing, but the standard deviation of their duration is decreasing. For $\omega>2.6$ we enter the synchronization region, where the value of $\rho_s$ drops due to the nearly consistently synchronized dynamics.

Finally, the mean phase velocities in the right column of Fig.\,\ref{fig.5} display the transition to frequency synchronization. While the frequency of the two driven nodes ($k=41,86$) converges to the frequency of an uncoupled FHN oscillator $\omega_{\text{FHN}} \approx 2.6$, also the frequencies of all the other nodes are adjusted, especially those with a much higher frequency ($k=18,63$).

\section{Conclusion}
We have investigated the influence of an external sound source on the dynamics of a network with empirical structural connectivity. It has been found that depending on the frequency and amplitude of the sound source, synchronization can be induced in the dynamics of the system. We have shown that two frequencies play an important role for synchronization, particularly the natural frequency of the uncoupled oscillator and the frequency of the coupled system. Moreover, the degree of synchronization is gradually increased when the frequency of the uncoupled oscillator or multiple values of it are approached. Furthermore, we have analyzed the linear dependence of the synchronization borders upon the amplitude of the external sound, which can also be characterized as the volume of the sound. This has resulted in the observation that the synchronization region can be enlarged by increasing volume. We have demonstrated the dynamical behavior of the system in the transition to synchronization. By tuning the frequency of the external sound appropriately, we have shown that the level of synchrony can be increased.

These results are in accordance with experiments of Bader's group~\citep{HAR14, HAR20a} that music induces a certain degree of synchrony in the human brain. This group has shown that listening to music can have remarkable influence on the brain dynamics, in particular, a periodic alternation between synchronization and desynchronization. Moreover, such an alternation reflects the variability of the system; this can be seen as a critical state between a fully synchronized and a desynchronized state. It is known that the brain is operating in a critical state at the edge of different dynamical regimes~\citep{MAS15a}, exhibiting hysteresis and avalanche phenomena as seen in critical phenomena and phase transitions \citep{KIM18,RIB10,STE10a}. By choosing appropriate parameters, we have reported an intriguing dynamical behavior regarding the transition to synchronization, and have observed the induced alternation between high and low degrees of synchronization. To sum up, an external sound source connected to the brain allows for synchronization dynamics, which may be used to model the effect of music on the human brain.

\section*{Conflict of Interest Statement}

The authors declare that the research was conducted in the absence of any commercial or financial relationships that could be construed as a potential conflict of interest.

\section*{Author Contributions}

JS did the numerical simulations and the theoretical analysis. ES supervised the study. All authors designed the study and contributed to the preparation of the manuscript. All the authors have read and approved the final manuscript.

\section*{Funding}
This work was supported by the Deutsche Forschungsgemeinschaft (DFG, German Research Foundation, project No. 429685422).

\section*{Acknowledgments}
We are grateful to Anton\'{i}n \v{S}koch and Jaroslav Hlinka for preparing the example structural connectivity matrices, and to Rolf Bader and Lenz Hartmann for stimulating discussions.

%

\bibliographystyle{frontiersinSCNS_ENG_HUMS} 



\end{document}